\documentclass[conference]{IEEEtran}
\IEEEoverridecommandlockouts

\usepackage{cite}
\ifCLASSINFOpdf
\usepackage[pdftex]{graphicx}
\else
\fi

\usepackage{amsfonts}
\usepackage{amsmath}
\usepackage{amssymb}
\usepackage{bbm}
\usepackage{xcolor}
\newcommand{\RNum}[1]{\uppercase\expandafter{\romannumeral #1\relax}}

\usepackage{epsfig}
\usepackage{epstopdf}
\usepackage{algorithm}
\usepackage{algpseudocode}
\usepackage{graphicx}
\usepackage{float}
\usepackage{subfig}
\usepackage{upgreek}

\ifodd 1
\newcommand{\rev}[1]{{\color{red}#1}} 
\newcommand{\com}[1]{\textbf{\color{blue} (Revised: #1)}}
\else
\newcommand{\rev}[1]{#1}
\newcommand{\com}[1]{}
\fi

\usepackage{enumitem}
\usepackage[font=small]{caption}
\makeatletter
\newcommand\tinyv{\@setfontsize\tinyv{7pt}{9}}
\makeatother

\usepackage{varwidth}

\begin{document}
\bibliographystyle{IEEEtran}
\bstctlcite{IEEEexample:BSTcontrol}
	
\title{Broadband Digital Over-the-Air Computation for Asynchronous Federated Edge Learning}
\author{\IEEEauthorblockN{Xinbo Zhao$^{\dag}$,
		Lizhao You$^{\dag *}$, 
		Rui Cao$^{\dag}$,
		Yulin Shao$^{\S}$,
		and Liqun Fu$^{\dag}$}
	\IEEEauthorblockA{$^{\dag}$School of Informatics, Xiamen University \\
	$^{\S}$Department of Electrical and Electronic Engineering, Imperial College
	London}
		Emails: \{xbzhao,~ruicao\}@stu.xmu.edu.cn, y.shao@imperial.ac.uk, \{lizhaoyou,~liqun\}@xmu.edu.cn
\IEEEcompsocitemizethanks{$^{*}$Corresponding Author: Lizhao You}	
}
	
\maketitle
\begin{abstract}
	This paper presents the first broadband digital over-the-air computation (AirComp) system for phase asynchronous OFDM-based federated edge learning systems. Existing analog AirComp systems often assume perfect phase alignment via channel precoding and utilize uncoded analog modulation for model aggregation. In contrast, our digital AirComp system leverages digital modulation and channel codes to overcome phase asynchrony, thereby achieving accurate model aggregation in the asynchronous multi-user OFDM systems. To realize a digital AirComp system, we propose a non-orthogonal multiple access protocol that allows simultaneous transmissions from multiple edge devices, and present a joint channel decoding and aggregation (Jt-CDA) decoder (i.e., full-state joint decoder). To reduce the computation complexity, we further present a reduced-complexity Jt-CDA decoder (i.e., reduced-state joint decoder), and its arithmetic sum bit error rate performance is similar to that of the full-state joint decoder for most signal-to-noise ratio (SNR) regimes. Simulation results on test accuracy (of CIFAR10 dataset) versus SNR show that: 1) analog AirComp systems are sensitive to phase asynchrony under practical setup, and the test accuracy performance exhibits an error floor even at high SNR regime; 2) our digital AirComp system outperforms an analog AirComp system by at least 1.5 times when SNR$\geq$9dB, demonstrating the advantage of digital AirComp in asynchronous multi-user OFDM systems.
\end{abstract}


\section{Introduction}

\begin{figure}[t!]
	\centering
	\setlength{\abovecaptionskip}{0.cm}
	\includegraphics [scale=0.275]{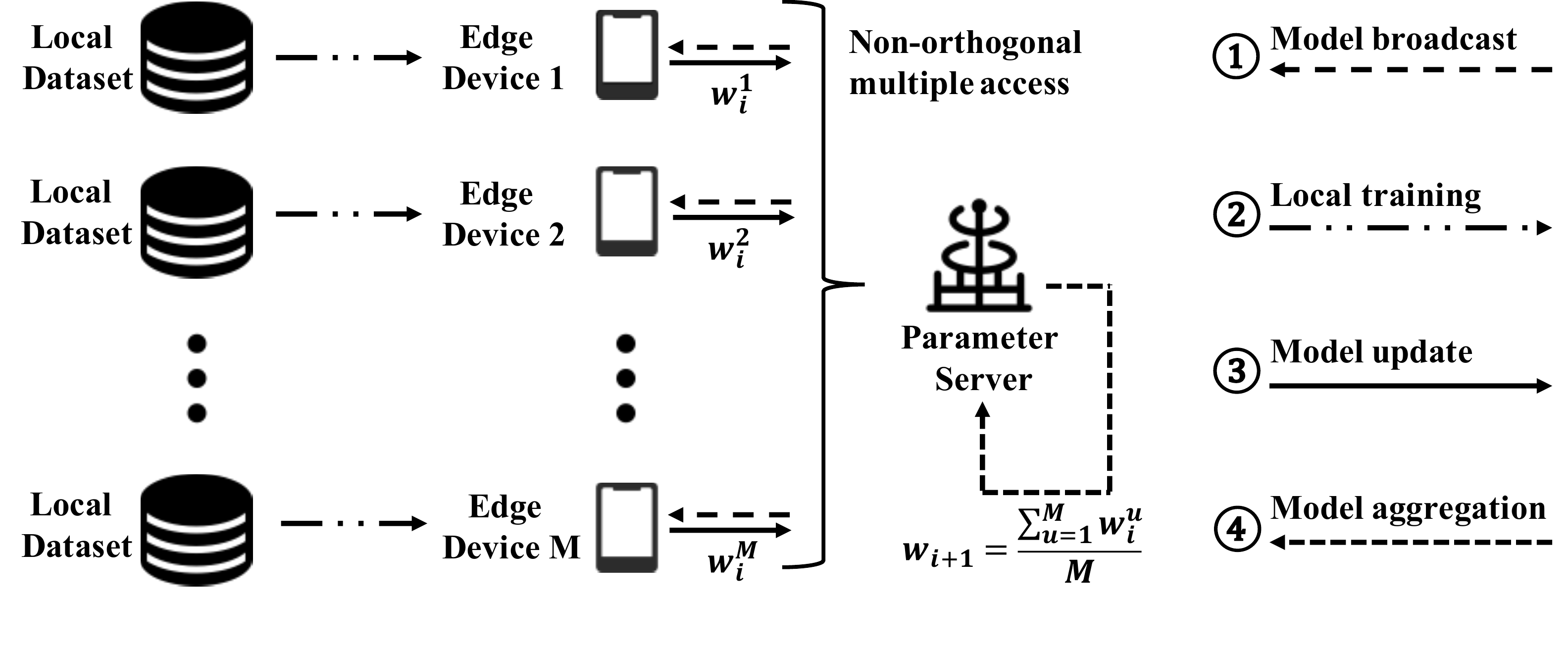}
	\caption{Federated edge learning via over-the-air computation.\vspace{-2ex}}\label{FL}
	\vspace{-2ex}
\end{figure}

Due to the increased concern in data privacy preserving, federated edge learning (FEEL) has become a popular distributed learning framework for mobile edge devices. A FEEL system often consists of a parameter server (PS) and multiple edge devices with non-independent and identically distributed (i.i.d.) local datasets, as shown in Fig.~\ref{FL}. PS periodically aggregates (sum and average) the locally-trained deep neural network models over the wireless channel.

Since large numbers of model parameters are transmitted in the model aggregation step, communication efficiency has become a bottleneck that restricts the development of FEEL \cite{McMahan2017}. Over-the-air computation (AirComp) has been proposed to address this problem\cite{amiri2019over,zhu2019broadband}. In AirComp systems, selected edge devices transmit simultaneously using non-orthogonal wireless resource, and the parameter server decodes an arithmetic sum of edge devices’ data directly from the superimposed signal. The non-orthogonal multiple access (NOMA) scheme plus aggregation directly at the physical-layer improves the communication efficiency, and thus accelerates FEEL. 

Existing AirComp systems \cite{amiri2019over, zhu2019broadband, xing2020decentralized, yang2020federated, amiri2020machine, zhu2020one, sery2021over, wang2021federated, shao2021denoising} are all analog AirComp systems that use analog modulation for communication. Specifically, the edge devices transmit their model parameters in an analog fashion without any channel coding. The overlapped signals in the air naturally produce the arithmetic sum of signals. The PS then extracts the arithmetic sum directly from the received and superimposed signal. However, the performance of analog AirComp is sensitive to the phase offsets among the superimposed signals: they can add up destructively if their signals are out of phase, resulting in large aggregation errors. Therefore, these analog AirComp systems assume perfect channel precoding that can compensate all channel impairments, such as path loss, fading, time synchronization errors, and carrier frequency offset (CFO), and receive clean arithmetic sum of model parameters. 

However, realizing such perfect phase-aligned transmissions is quite challenging in practice, especially for OFDM broadband systems. Considering a point-to-point OFDM system for example, a tiny time synchronization error turns into a linear phase increase over all subcarriers; a tiny residual CFO rotates subcarriers’ phases over symbols. For a multi-user OFDM system, different users may have different symbol arrival time and CFOs, leading to phase asynchrony in all subcarriers (i.e., different phase offsets over subcarriers and rotating phase offsets over symbols). 

To tackle this problem, this paper puts forth the first OFDM-based digital AirComp system. Unlike analog AirComp that hinges on accurate channel precoding, digital AirComp leverages digital modulation and channel codes to combat channel impairments and misalignments, thereby achieving accurate model aggregation even when the signal phases of multiple edge devices are misaligned at the PS. Specifically, in channel-coded digital AirComp system, model parameters are first quantized into bits. The source bits are then encoded by channel codes; modulated into constellations; OFDM modulated into channel symbols; and transmitted to the receiver.
Meanwhile, the receiver aims to decode the aggregated source bits instead of the individual source bits transmitted from different edge devices. This process is referred to as the channel decoding and aggregation (CDA) process.



Our findings and main contributions are as follows:

\begin{enumerate}
	\item We put forth a new digital AirComp system for OFDM-based broadband FEEL. Compared with analog AirComp systems, our digital AirComp system is robust to phase asynchrony among edge devices, and thus eliminates the need of channel precoding required by analog AirComp.
	\item We design two joint channel decoding and aggregation (Jt-CDA) decoders for convolutional-coded AirComp. The first Jt-CDA decoder, \emph{full-state} joint decoder (FSJD), is a nearly optimal decoder that leverages the log-max approximation to realize maximum likelihood (ML) decoding. The second Jt-CDA decoder, \emph{reduced-state} joint decoder (RSJD), is a simplified version of FSJD that greatly reduces the number of states in ML decoding without compromising much performance.
	\item We perform extensive simulations to evaluate the performance of the proposed digital AirComp system against the analog AirComp system in asynchronous OFDM-based FEEL systems. Simulations results on the test accuracy performance verify that the analog AirComp exhibits an error floor in the presence of phase misalignments; the proposed digital AirComp, in contrast, completely eliminates the error floor and outperforms the analog AirComp when the signal-to-noise ratio (SNR) is larger than 9 dB. 
\end{enumerate}

The remainder of this paper is organized as follows. Section \ref{related} overviews related work. Section \ref{sys mod} presents our system models. Section \ref{SDD} introduces the various proposed Jt-CDA decoders. Simulation results are showed in Section \ref{simu}. Section \ref{conclusion} concludes the paper.

\section{Related Work} \label{related}

Over-the-air computation in multiple-access channels is first studied from information-theoretic perspective \cite{gastpar2003source, nazer2007computation}. They show that the joint communication and computation can be more efficient than the separate communication from computation in terms of function computation rate. Recently, many AirComp systems \cite{amiri2019over, zhu2019broadband, xing2020decentralized, yang2020federated, amiri2020machine, zhu2020one, sery2021over, wang2021federated, shao2021denoising} for FEEL have been proposed. However, they are all analog AirComp systems, and require perfect synchronization on the symbol and phase level, which is hard to realize in practice, especially for OFDM systems. More specifically, although many distributed MU-MIMO systems \cite{balan2013airsync, abari2015airshare, hamed2018chorus} can realize oscillator synchronization (i.e., removing CFOs), totally phase-aligned transmissions are still quite challenging (there is an attempt in \cite{abari2016over} with some analysis, but the real system is not demonstrated).

Some analog AirComp systems are proposed to handle misaligned transmissions. Goldenbaum et al., \cite{goldenbaum2013robust, kortke2014analog} design and implement an analog-modulated AirComp system that only needs coarse symbol-level synchronization, and leverages direct spread spectrum sequences with random phases for modulation. Shao et al., \cite{shao2021federated,shao2021bayesian} leverages oversampling and sum-product ML estimators to estimate the arithmetic sum under symbol misalignment and phase misalignment. However, their designs only work for conventional time-domain narrowband systems, and cannot directly apply to OFDM-based frequency-domain broadband systems.

Our digital AirComp design is inspired by physical-layer network coding (PNC) \cite{zhang2006hot}. In PNC, the receiver is interested in obtaining finite field sum. Many decoders \cite{to2010convolutional,you2016reliable,ullah2017phase} have been proposed to deal with symbol asynchrony and phase asynchrony in multi-user OFDM systems. Instead, our receiver is interested in obtaining arithmetic sum, making the channel decoder design different.

\section{System Model} \label{sys mod}
In this section, we first introduce the FEEL architecture and our digital AirComp transceiver architecture. Then we introduce the asynchronous transmission model and the digital AirComp reception model, as well as an NOMA protocol.


\subsection{Federated Edge Learning Model}
FEEL aims to learn a joint model with the help of a number of edge devices without transmiting raw datasets. The joint model training requires many iterations to converge. Each iteration contains the following four steps:
\begin{enumerate}
\item Model broadcast: $M$ edge devices are randomly selected by the parameter server and each of them is broadcasted with the current global model $w_i$ in the $i$th iteration;
\item Local training: every edge device $u$ trains the received global model with their local data and gets a new local model $w_i^u$;
\item Model update: the local models of selected edge devices are transmitted to parameter server;
\item Model aggregation: the parameter server takes the average of the received models to update the global model for the next iteration $i+1$: $w_{i+1}=\sum_{u=1}^{M}w^u_i/M$.
\end{enumerate}

Due to the resource constraints of mobile edge devices, FEEL is supported by wireless networks (e.g., 5G or WiFi 6). Therefore, the channel access performance is crucial to the performance of FEEL.

\begin{figure}[t!]
	\centering
	\setlength{\abovecaptionskip}{2.mm}
	\includegraphics [scale=0.3,trim=0 0 0 0]{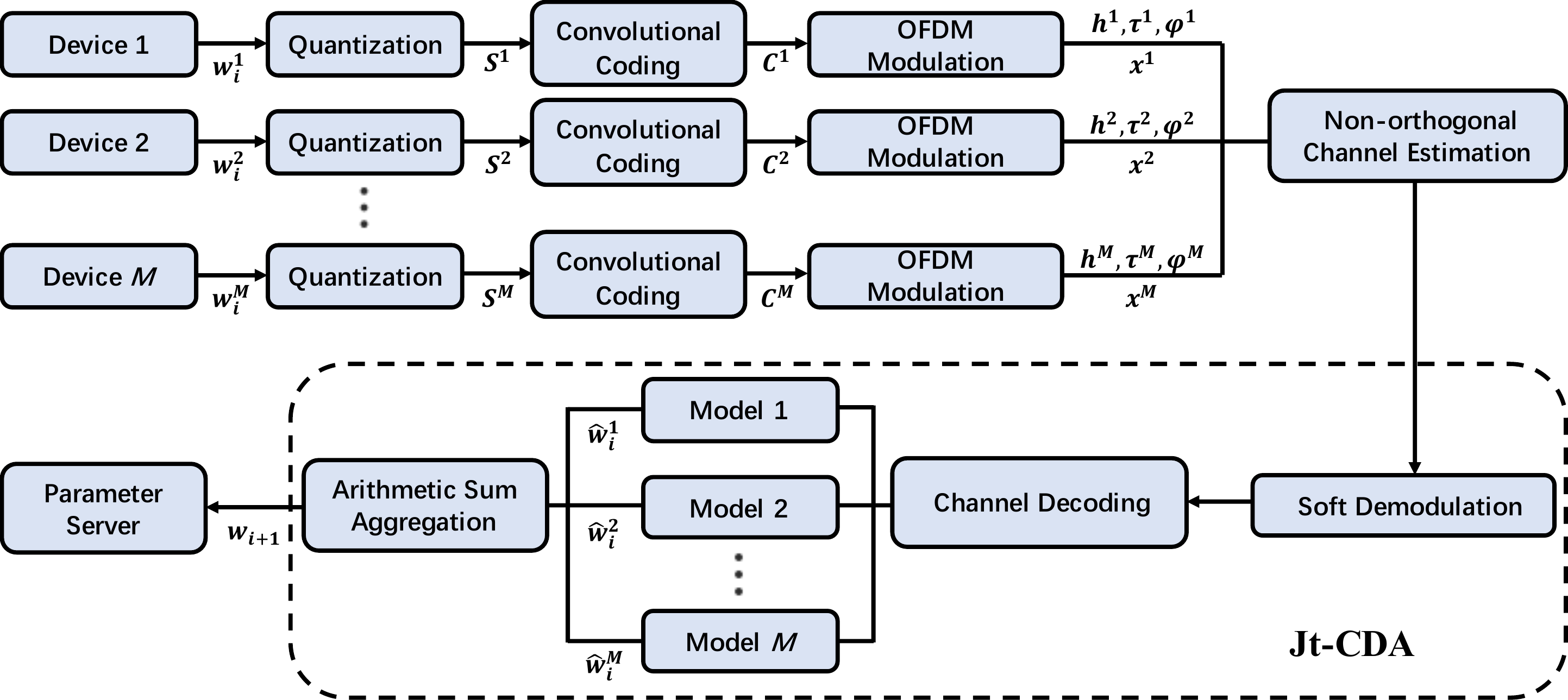}
	\caption{Broadband digital AirComp transceiver architecture.\vspace{-2ex}}\label{arch}
	\vspace{-2ex}
\end{figure}

\subsection{Broadband Digital AirComp Transceiver} 
The model update and aggregation steps require large bandwidth. 
Therefore, in this paper we consider OFDM wideband systems for model update, especially 802.11 OFDM systems. 
To improve spectrum efficiency, we focus on NOMA that supports model update and aggregation simultaneously. 

Fig.~\ref{arch} shows our broadband digital AirComp transceiver architecture. To be best compatible with the 802.11 standard, our system adopts most components as in conventional 802.11 OFDM systems and makes minimal modifications. In particular, the transmitter architecture is almost the same with the traditional OFDM transmitter except that we allocate orthogonal pilots to each transmission edge devices for channel estimation. For the receiver, we follow the conventional frame detection and multi-user channel estimation, but redesign the demodulation and channel decoding for the arithmetic sum computation. 

\subsection{Asynchronous Communication Model}  \label{AAirComp}



Our system first quantizes the updated model parameter that each edge device transmits with quantization bit length $k$. Let $p$ be one of the transmitted floating-point value parameters of local model $w$, and $s$ be the quantized binary bit sequence. The quantization step can be represented as:

\begin{small}
\begin{equation}
	s_0=\left\{
		\begin{array}{lr}
		1,\!\!& p < 0 \\
		0,\!\!& p\ge 0
		\end{array}
	\right. \!\!; 
	s_i=\left\{
		\begin{array}{lr}
		1,\!\!& |p|\bmod \frac{1}{2}^{i-1}\ge{\frac{1}{2}}^i \\
		0,\!\!& |p|\bmod \frac{1}{2}^{i-1}<{\frac{1}{2}}^i 
		\end{array}
	\right. \!\!,
	i\ge 1.
	\vspace{-1ex}
\end{equation}
\end{small}

Then we can pack several parameters in a packet for transmisson (assuming an OFDM packet can transmit $n$ source bits, we can pack $l = \lceil n / k \rceil$ parameters). Let $S=\{s_0, \dots, s_{n-1} \}$ be the source bits fed to the OFDM transmitter. After channel encoding (e.g., convolutional codes), the coded bits $C$ are modulated in the frequence domain. We assume that BPSK is used in this paper, and each bit $C[i]$ is modulated to $X[i]$ with $X[i] \in \{+1,-1\}$. Then after Inverse Discrete Fourier Transform (IDFT), adding Cyclic Prefix (CP) and preambles, and Digital-to-Analog convertion (DAC), $X$ is transformed to time-domain continuous signal $x(t)$.



In the following, we use two edge devices (devices $A$ and $B$) that are selected transmit simultaneously for example (i.e., $M=2$), and show how the parameter server computes the aggregation data from the superimposed signal. Extension to $M\geq2$ is straightforward.

\textbf{Asynchronous Multi-User Transmission:}
Let $x^u(t)$ be the time-domain signal corresponding to $X^u$ ($u \in \{A, B\}$). Then, the received baseband signal $y(t)$ in parameter server can be represented as:
\begin{equation}
	y(t) = \sum_{u \in \{A, B\}} (h^u(t) * x^u(t-\tau^u)) e^{j \phi^u(t)} + n(t),
	\vspace{-1ex}
\end{equation}
where $h^u(t)$ is the channel impulse response of the edge device $u$, $*$ is the convolution operation, $\tau^u$ of device $u$ is the time offset with respect to the start of receiving window in the parameter server, $e^{j\phi^u(t)}$ is the phase offset caused by the carrier frequency offset (CFO) with respect to the parameter server, and $n(t)$ is the zero-mean circularly-symmetric complex additive white Gaussian noise (AWGN). 

Note that $\tau^u$ and $\phi^u(t)$ exist in practical system due to time synchronization error and the intrinsic CFO. In commercial multi-user OFDM systems (e.g., UL MU-MIMO), if the relative time offset $|\tau^A-\tau^B|$ is less than the CP duration, the receiver can find a proper CP cut for decoding and transform time offset to subcarriers' phase offsets. Although many papers assume a perfect phase precoding scheme that can compensate all time and frequency errors, it is quite challenging for implementation. Therefore, we consider the above asynchronous simultaneous transmission model.


\textbf{Reception and Aggregation:}
Let $k$ denotes the symbol index, and $n$ denotes the subcarrier index. The received baseband signal of the $k$-th symbol and the $n$-th subcarrier in frequency domain can be represented as:
\begin{equation}
	Y_k[n]= H_k^A[n]X_k^A[n]+H_k^B[n]X_k^B[n]+N_k[n],
\end{equation}
where $H_k^A[n]$ and $H_k^B[n]$ are the frequency-domain channel of the two edge devices, $X_k^A[n]$ and $X_k^B[n]$ are modulated signal of the two edge devices. Our target is to compute the arithmetic sum $S^F=S^A+S^B=\{s_0^A+s_0^B, \cdots, s_{n-1}^A+s_{n-1}^B\}$ of all parameters from all data subcarriers' signal, where $+$ denotes the arithmetic sum.

Here $H_k^u[n]$ is the combination effect of wireless channel, time offsets and CFOs. In particular, the tiny time offset between the transmitter and the receiver corresponds to a linear phase increase over all subcarriers (which is the same for all symbols), and CFO leads to phase accumulation over all symbols for each subcarrier. Since the time offset and CFO are different for devices $A$ and $B$, the phase difference of their channel $H_k^u[n]$ are not constant over all symbols, making digital aggregation decoding challenging.




\subsection{Non-Orthogonal Multiple Access Protocol} \label{protocol}

\begin{figure}[t!]
	\centering
	\setlength{\abovecaptionskip}{0.cm}
	\includegraphics [scale=0.265,trim=0 0 0 0]{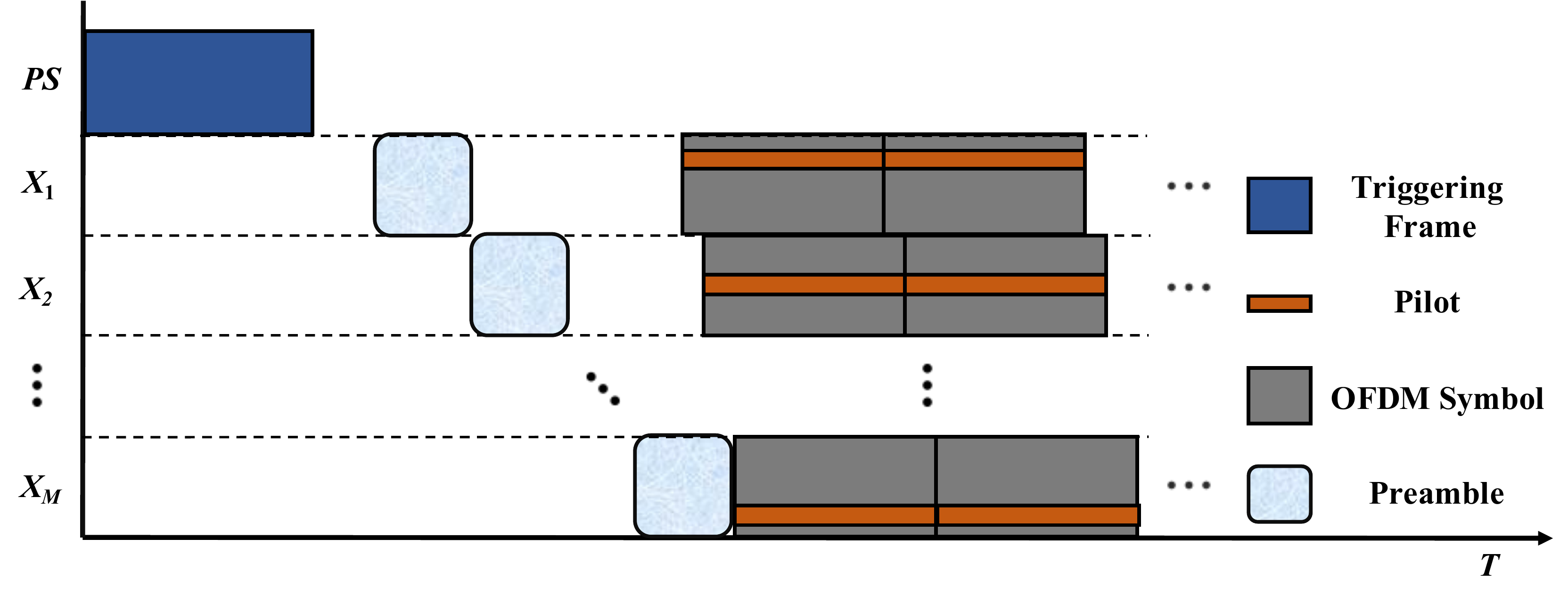}
	\caption{Multi-user non-orthogonal multiple access protocol.}
	\label{nnma}
	\vspace{-3ex}	
\end{figure}

Fig.~\ref{nnma} shows the proposed NOMA protocol. Similar to the multi-user simultaneous transmission protocol in 802.11ax (e.g., UL-MIMO), the parameter server first transmits a triggering frame that notifies the transmission edge devices in this round. The informed edge devices synchronize to the triggering frame, and then transmit OFDM frames simultaneously with their preambles orthogonal in time-domain and data symbols overlapped. Moreover, we allocate orthogonal pilots to each edge device in all OFDM symbols. In this way, receiver can estimate initial frequency-domain channel through orthogonal preambles, and track phase rotation through orthogonal pilots.

\section{Channel Decoding and Aggregation Design} \label{SDD} 
This Section presents the design of some Jt-CDA decoders for digital AirComp systems with convolutional codes. 
Before we present technical details, we first introduce the ``codeword'' optimal Jt-CDA decoder. For devices $A$ and $B$, their source packets are $S^A$ and $S^B$ with length $K$. Let the encoding function be $\Pi(\cdot)$. Their encoded packets (codewords) are $C^A=\Pi(S^A)$ and $C^B=\Pi(S^B)$ with length $N$. Their modulated signals are denoted by $X^A$ and $X^B$. Let $Y$ denote the received signal. Our target is to get the arithmetic sum of source packets $S^F = S^A + S^B$. Note that there is no valid codeword corresponding to $S^F$ since $+$ operation is not defined in $\Pi(\cdot)$.

More specifically, our target is to compute 

\begin{small}
\begin{equation} \label{eq1}
	\vspace{-1ex}
	\begin{split}
		\hat{S}^{F} &= \operatorname*{arg\,max}_{S^{F}} \sum_{C^A, C^B: \Pi^{-1}(C^A)+\Pi^{-1}(C^B)=S^F} Pr(Y|C^A,C^B) \\
		&= \operatorname*{arg\,max}_{S^{F}} {log} \sum_{C^A, C^B: \Pi^{-1}(C^A)+\Pi^{-1}(C^B)=S^F} Pr(Y|C^A,C^B),
	\end{split}
	\vspace{-1ex}
\end{equation}
\end{small}
where
\begin{small}
\begin{equation} \label{eq2}
	\vspace{-1ex}
	\begin{split}
		&Pr(Y|C^A,C^B)=\prod_{n=1}^N Pr(Y[n]|C^A[n],C^B[n]) \\
		&= \prod_{n=1}^N exp(-\frac{|Y[n]-H^A[n]X^A[n]-H^B[n]X^B[n]|^2}{2\sigma^2}) \\
		&= exp(\sum_{n=1}^N -\frac{|Y[n]-H^A[n]X^A[n]-H^B[n]X^B[n]|^2}{2\sigma^2}),
	\end{split}
	\vspace{-1ex}
\end{equation}
\end{small}
Equation (\ref{eq1}) means that we need to add up the probability of all possible codeword pairs ($C^A$, $C^B$) (and  corresponding $S^A$, $S^B$) that generate the same $S^F$. In fact, it is a $4^K$-to-$3^K$ mapping, since $\{0,1\} + \{0,1\} \rightarrow \{0,1,2\}$. There is no known exact computation method for Equation (\ref{eq1}) except to exhausively sum over all possible combinations, which has prohibitively high computation complexity. Since we consider a complete packet $S^F$, we still call it ``codeword''-optimal decoder, although $S^F$ does not correspond to a valid codeword.

\subsection{Full-State Joint Decoder (FSJD)}\label{FSJD}

Full-state joint Viterbi decoder (FSJD) is an approximated decoder for the Maximum Likelihood (ML) decoder. In particular, Equation (\ref{eq1}) can be simplified using log-max approximation (i.e., $log(\sum_i exp(x_i)) \approx \max_i x_i$) to 

\begin{small}
\begin{equation} \label{eq3}
	\begin{split}
		\hat{S}^{F} &\approx \operatorname*{arg\,max}_{S^{F}} \max_{C^A, C^B: \Pi^{-1}(C^A)+\Pi^{-1}(C^B)=S^F} Pr(Y|C^A,C^B) \\
		&\approx \operatorname*{arg\,min}_{S^{F}} \min_{C^A, C^B: \Pi^{-1}(C^A)+\Pi^{-1}(C^B)=S^F}  \frac{\Lambda(X^A,X^B)}{2 \sigma^2} \\
		&\approx \operatorname*{arg\,min}_{S^{F}} \min_{C^A, C^B: \Pi^{-1}(C^A)+\Pi^{-1}(C^B)=S^F}  \Lambda(X^A,X^B),
	\end{split}
	\vspace{-2ex}
\end{equation}
\end{small}
where
\begin{small}
\begin{equation} \label{eq4}
	\Lambda(X^A,X^B)=\sum_{n=1}^N (|Y[n]-H^A[n]X^A[n]-H^B[n]X^B[n]|)^2,
	\vspace{-2ex}
\end{equation}
\end{small}

The computation of Equation (\ref{eq3}) includes two steps: 1) find the best pair of codewords $\hat{C}^A$ and $\hat{C}^B$ such that $(\hat{C}^A,\hat{C}^B) = \operatorname*{arg\,min}_{C^A,C^B} \Lambda(X^A,X^B)$; 2) map $\hat{C}^A$ to $\hat{S}^A$ and $\hat{C}^B$ to $\hat{S}^B$, and get $\hat{S}^F = \hat{S}^A + \hat{S}^B$. Step 1) is equivalent to finding the minimum-cost path on the joint trellis of device A's and device B's encoders, and can be implemented by using the Viterbi algorithm. In Equation (\ref{eq4}), for BPSK-modulated received signal $Y[n]$, we can construct four possible constellations with $H^A[n]$ and $H^B[n]$ (since $X^A[n],X^B[n] \in \{+1,-1\}$), and calculate its Euclidean distance to the received point as edge cost. Since the state space of the joint trellis is the combination of device A's and device B's state space, we call it \emph{full-state} joint Viterbi decoder.

For conventional single-user trellis of convolutional codes with constraint length $L$, the number of state is $2^{(L-1)}$. Each state branches out two edges to two states in next stage, and each edge is associated with $1$ input bit and $r$ output bits for a convolution code with coding rate $r$. For the two-user joint trellis case, the number of states is $2^{2(L-1)}$, and each state branches out four edges to four states in next stage. Moreover, each edge is associated with $2$ input bits (each device owns $1$ bit) and $2r$ output bits (each device owns $r$ bits).

The Viterbi decoding algorithm is to find the minimum-cost path from the start state to the end state. In particular, for each decoding state, it performs the Add-Compare-Select (ACS) operation to compute and choose the minimum-cost path to the current state (i.e., path metric), and records the previous best state for back-tracing. The path metric value of the $k$-th state at the $i+1$-th decoding stage can be represented as:
\begin{equation}\label{PM}
	PM[k,i+1] =\underset{k'\in{set(k)}}{min}\left(PM[k',i]+BM[k'\rightarrow{k}]\right),
\end{equation}
where $set(k)$ is the set of four states that branch out to state $k$ at the last stage $i$, and $BM[k' \rightarrow k]$ is the branch metric from state $k'$ to $k$. Note that $BM[k' \rightarrow k]$ is the sum of correpsonding bits' Euclidean distance in the constellation graph, as shown in Equation (\ref{eq4}).

After running the algorithm to the last decoding stage, we can determine the end state and identify the minimum-cost path from the start state to the end state. By tracking back the path, we can recover the joint source bits corresponding to the minimum-cost path and obtain the sum bits by summing the decoded source bits. The sum bits are then transformed back to floating-point values corresponding to each parameter, and average is performed to get the update model parameter value.

\rev{
}

\subsection{Reduced-State Joint Decoder (RSJD)}
The computation complexity of FSJD is proportional to the number of states in each decoding stage, since each state is associated with one ACS operation. The aforementioned FSJD has a high computation complexity, since there are $2^{2(L-1)}$ states in each decoding stage. For the convolution codes defined in 802.11, the constaint length is 7, and there are 4096 states. The state number increases exponentially as the user number $n$ increases (e.g., $2^{n(L-1)}$), leading to a high computation cost. 

Reduced-state joint Viterbi decoder (RSJD) is a simplified (or approximated) version of FSJD. In particular, it limits the number of states participating in the ACS operation in the next decoding stage. At most $R$ states with minimum path metric are selected to advance to the next decoding stage, and other states are ignored. Therefore, we call it \emph{reduced-state} joint Viterbi decoder.

In our joint trellis, $R$ states branch out to at most 4$R$ states in the next decoding stage, but RSJD still selects at most $R$ states for the following decoding stages. We use the quick-sort algorithm to select these states in our implementation. In this way, RSJD reduces the computation complexity, although RSJD may not be the optimal in terms of the minimum-cost path, since some paths get lost and are ignored. However, the performance does not degrade for most cases. In Section \ref{simu}, we will show that the performance of RSJD approaches that of FSJD for most SNRs, especially for high SNR regime.


\begin{figure*}[t!]
	\centering
	\subfloat[AWGN with perfectly aligned channel phase.]{
		\includegraphics[width=2.25in]{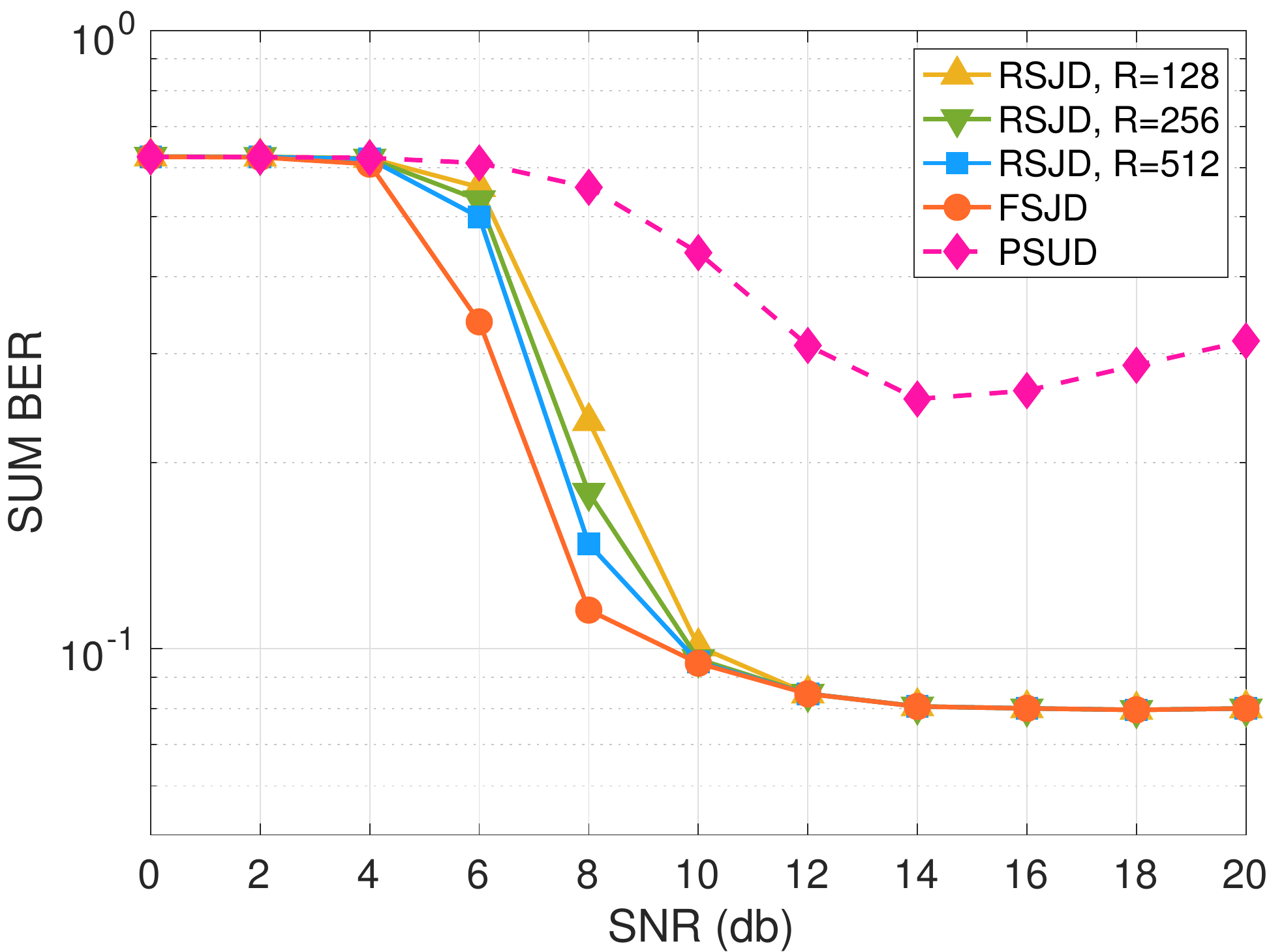} 
		\label{fig:phase0}}
	\subfloat[AWGN with perfectly orthogonal channel phase.]{
		\includegraphics[width=2.25in]{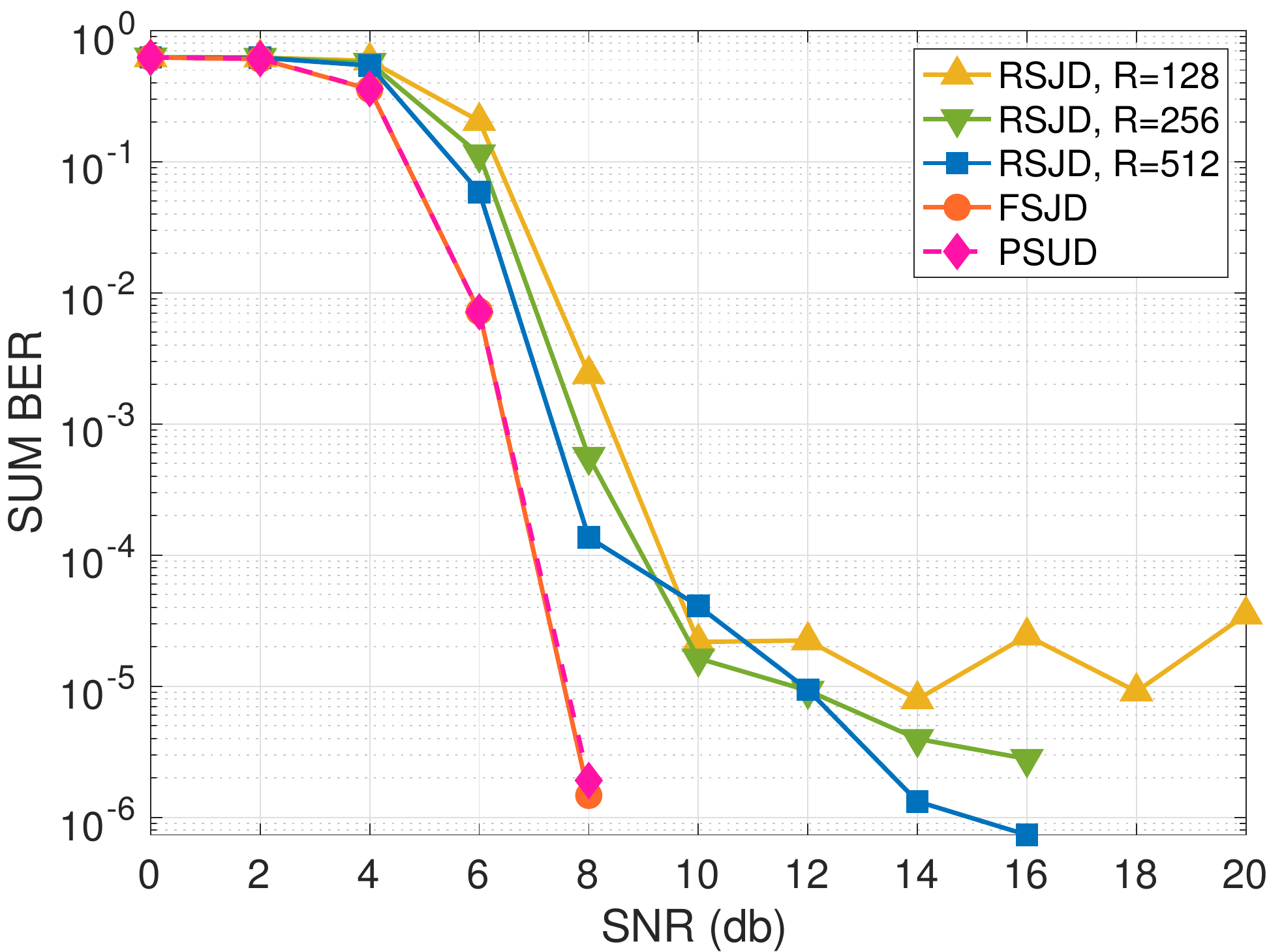} 
		\label{fig:phase90}}
	\subfloat[AWGN with random misaligned channel phase.]{
		\includegraphics[width=2.25in]{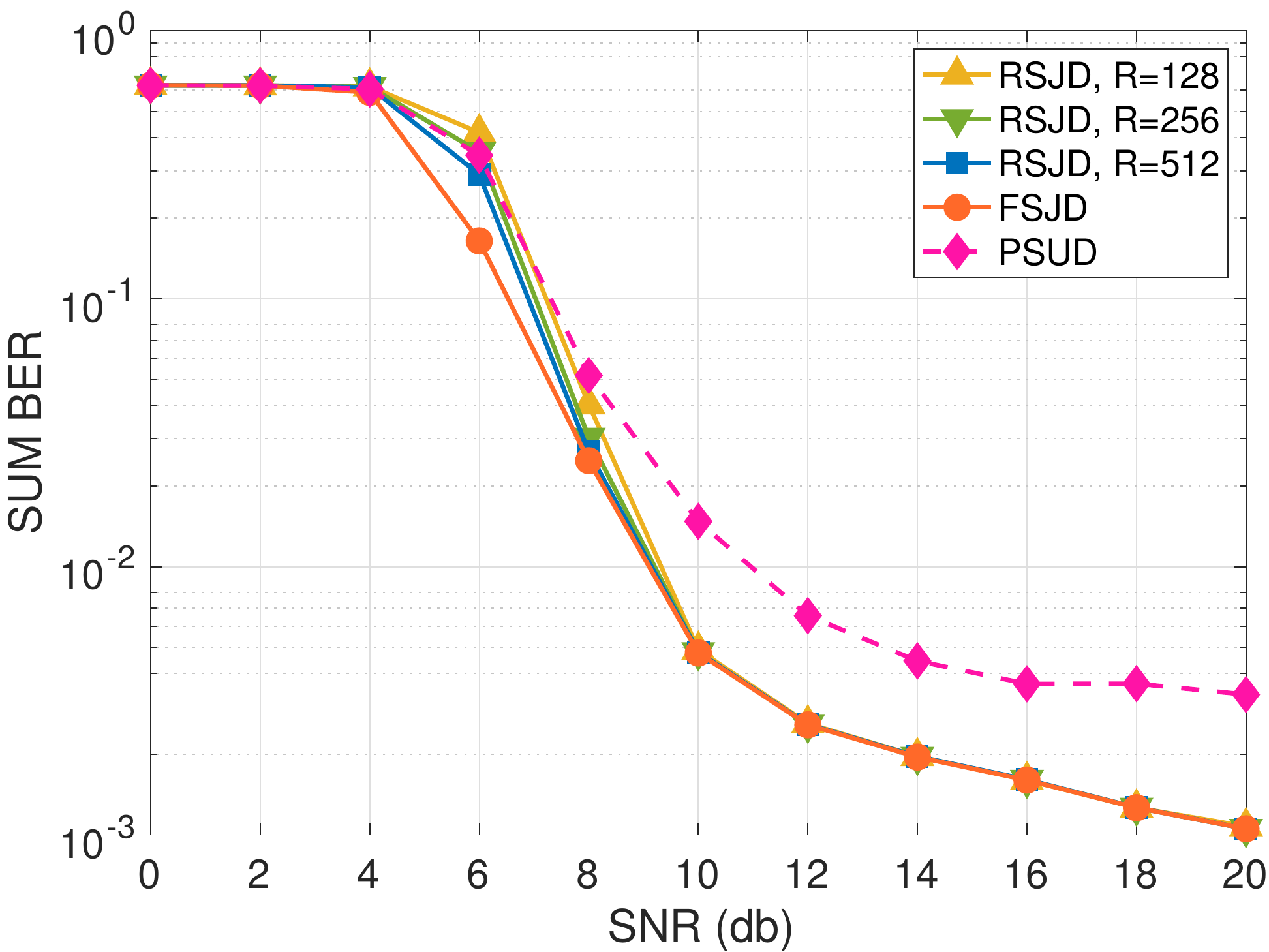} 
		\label{fig:phase_misalign}}
	\caption{SUM BER versus SNR at different channel settings.}
	\label{fig:BER_SNR}
	\vspace{-2ex}
\end{figure*}

\subsection{Parallel Single-User Decoders (PSUD)}
Parallel single-user decoder (PSUD) is another type of channel decoding and aggregation approach that leverages multi-user decoding (MUD) techniques. Different from joint decoders, PSUDs decode each device's data separately. In particular, we first decouples the superimposed signal in each subcarrier to each device's symbol decoding likelihoods
\begin{equation}
	Pr(Y[n]|C^A[n]) = \sum_{C^B[n]} Pr(Y[n]|C^A[n],C^B[n]),
	\vspace{-1ex}
\end{equation}
and
\begin{equation}
	Pr(Y[n]|C^B[n]) = \sum_{C^A[n]} Pr(Y[n]|C^A[n],C^B[n]),
	\vspace{-1ex}
\end{equation}
Then the decoupled symbol likelihoods are fed into two conventional single-user Viterbi decoder to find the codeword-optimal codewords ($\hat{C}^u$, $u \in \{A, B\}$) and source bits ($\hat{S}^u$) of two devices respectively. That is,
\begin{equation}
	\hat{C}^u = \operatorname*{arg\,max}_{C^{u}} \prod_{n=1}^N Pr(Y[n]|C^u[n]),
\end{equation}

Compared with FSJD, PSUD also reduces complexity, since PSUDs are two single-user decoders. The complexity of RSJD may still be higher than that of PSUD, depending on the number of kept states. However, the performance of PSUD is worse than RSJD and FSJD, as shown in Section \ref{simu}.

\section{Simulation Results}\label{simu}
In this section, we present simulation results on decoding performance of the presented Jt-CDA decoders, and the FEEL system performance based on digital AirComp.

\subsection{Federated Learning Setup}

Our FEEL system consists of one parameter server and 40 edge devices. They communicate through the wireless medium according to the NOMA protocol presented in Section \ref{protocol}. Each edge device uses a convolution neural network (CNN) called ShuffleNet V2 network\cite{Ma2018} to perform classification on CIFAR-10 dataset\cite{Krizhevsky2009}. The CIFAR-10 contains $32\times32$ RGB color images in 10 categories. The size of CIFAR-10 training dataset is $50,000$ and the size of testing dataset is $10,000$. The training datasets are distributed in a non-i.i.d. manner: 1) we first equally distribute the 40000 training images to 40 edge devices in random order; 2) the rest 10000  images are sorted by labels and equally distributed to every device with size 250 (minic non-i.i.d. data distribution). 

Each ShuffleNet V2 network has $1.26\times10^6$ parameters in total. We randomly choose $N=2$ edge devices for training in each iteration and update the model parameters based on the provided global model and local datasets every 5 epochs. Floating point parameters are quantized to bit sequences by the model presented in Section \ref{AAirComp}.

\begin{figure*}[!h]
	\begin{minipage}[t]{0.335\linewidth}
		\centering
		\setcaptionwidth{2.1in}
		\includegraphics[scale=0.29]{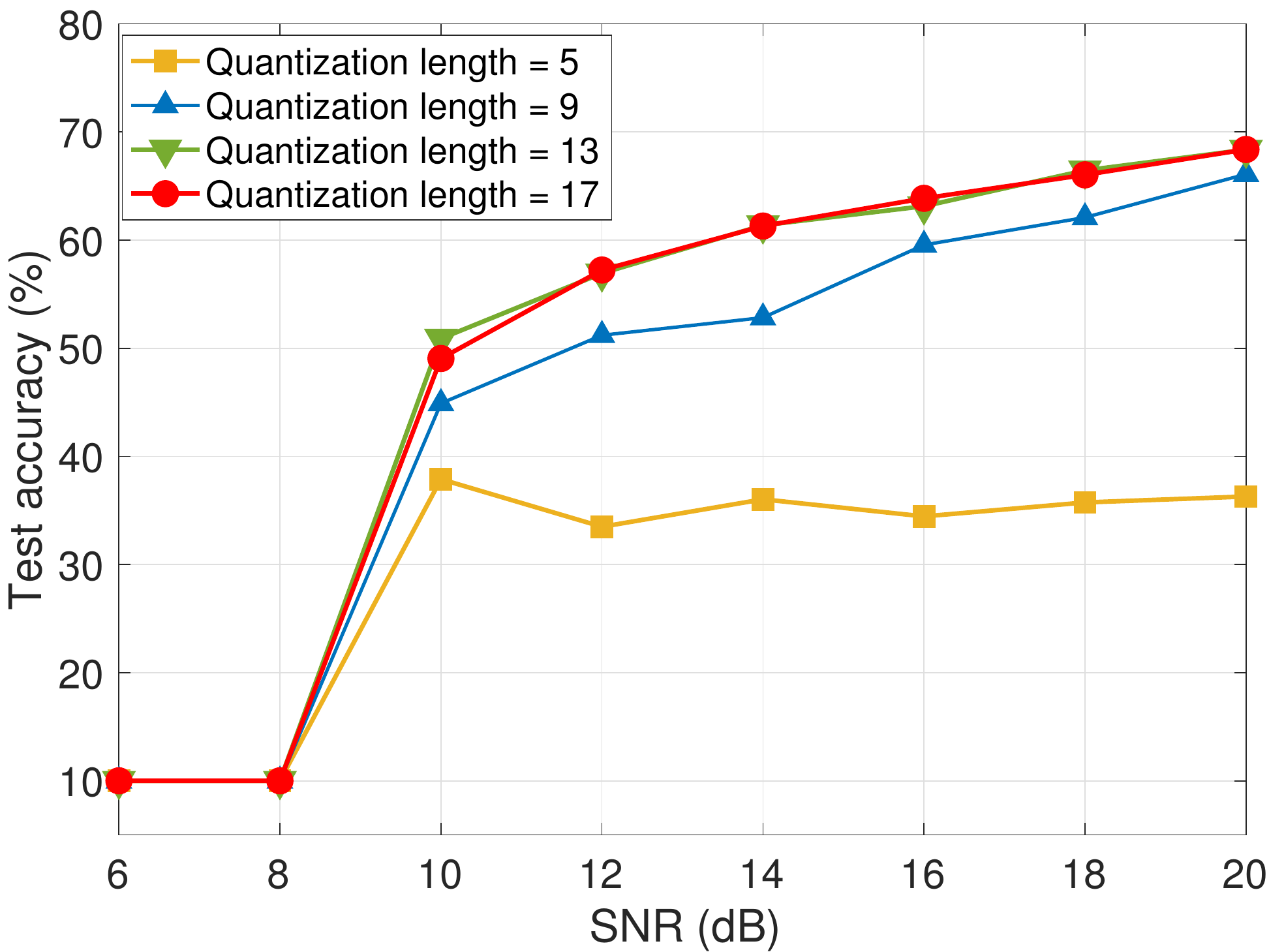} 
		\caption{Test accuracy versus SNR with different quantization lengths (digital AirComp).\vspace{-2ex}}
		\label{quant}
	\end{minipage}%
	\begin{minipage}[t]{0.335\linewidth}
		\centering
		\setcaptionwidth{2.1in}
		\includegraphics[scale=0.29]{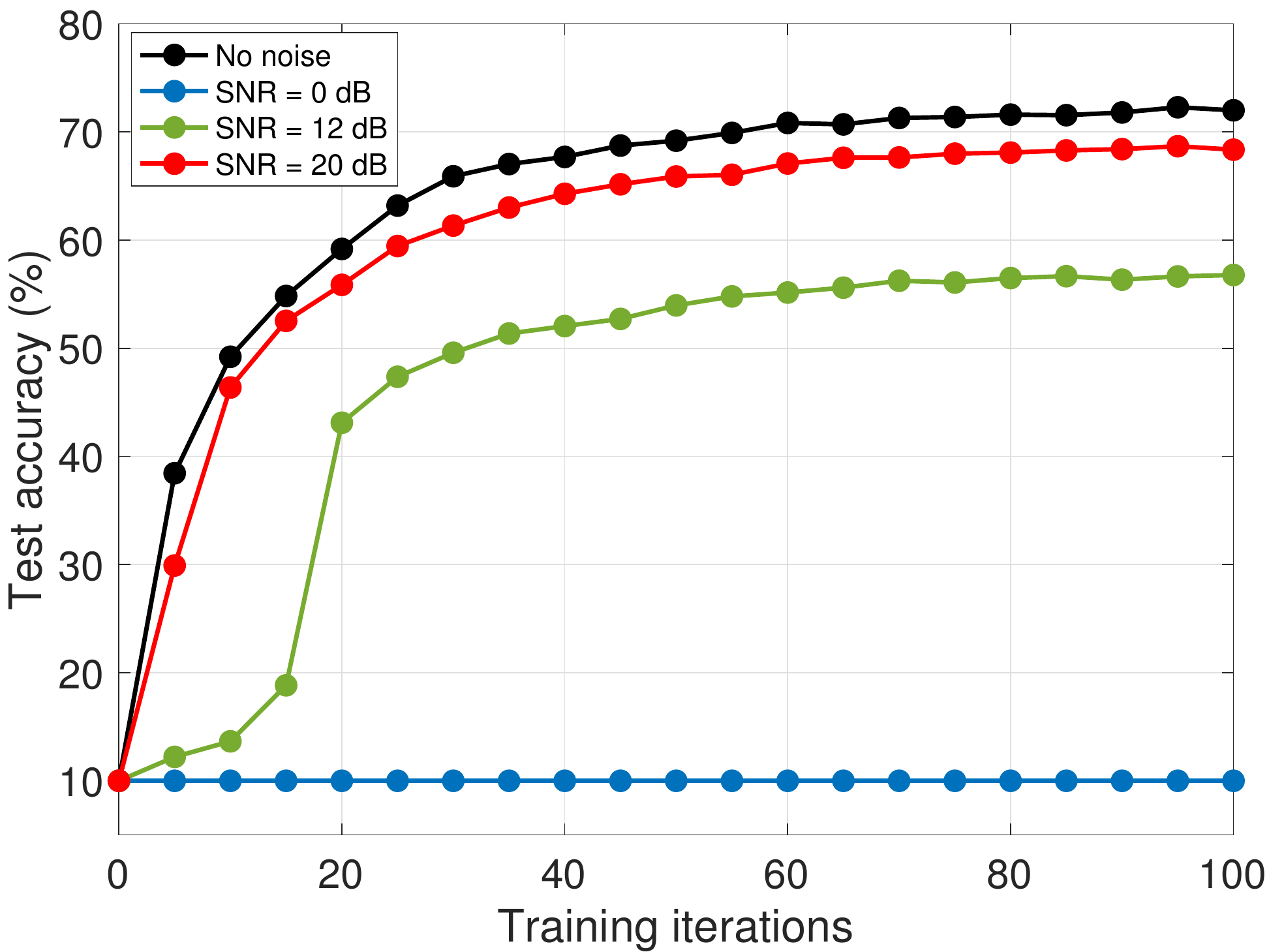} 
		\caption{Test accuracy over training iterations with different SNRs (digital AirComp).\vspace{-2ex}} 
		\label{diff_noise}
	\end{minipage}%
	\begin{minipage}[t]{0.335\linewidth}
		\centering
		\setcaptionwidth{2.1in}
		\includegraphics[scale=0.29]{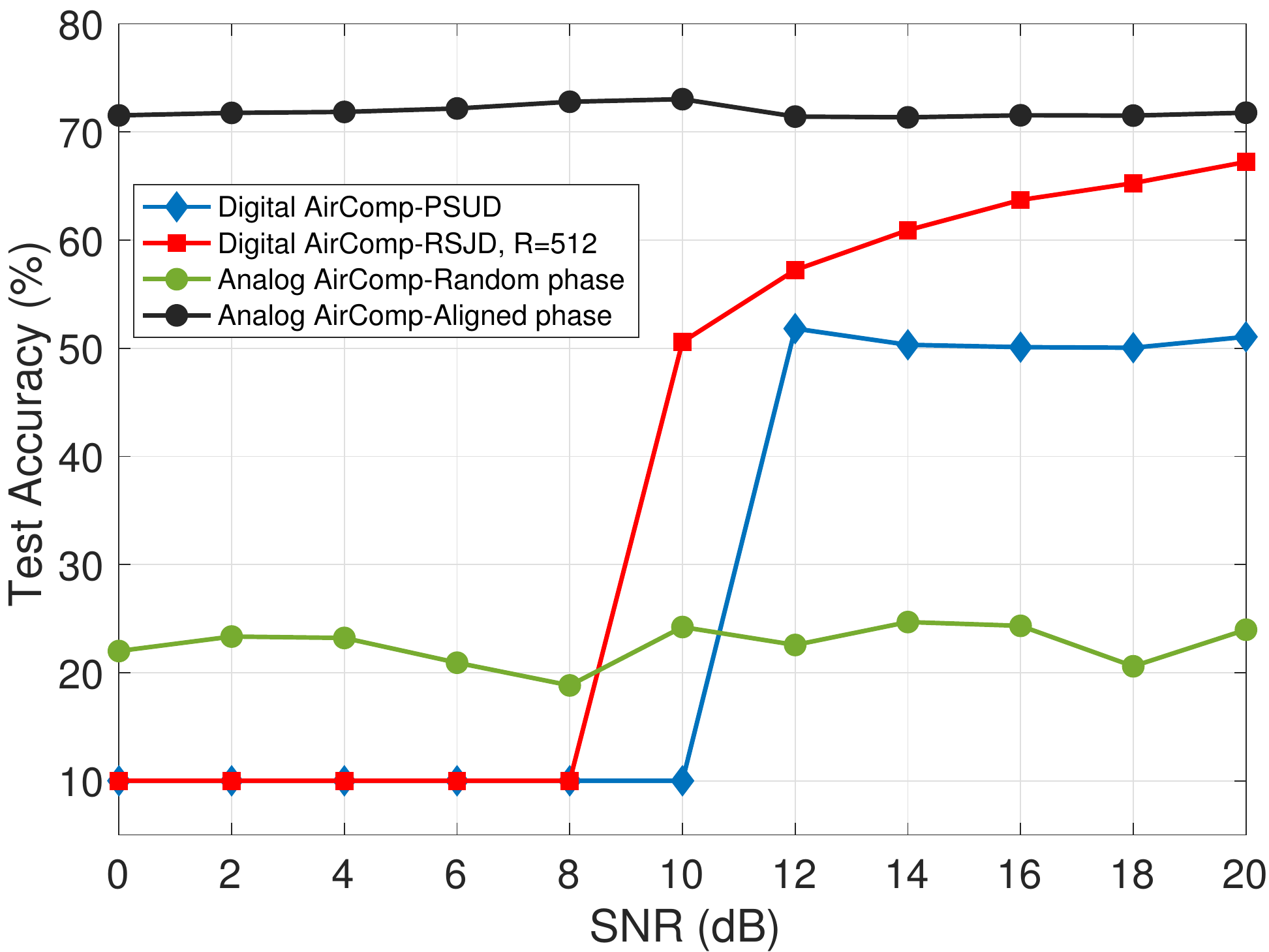} 
		\caption{Test accuracy of different AirComp methods under near-realistic channel.}
		\label{real-simu}
	\end{minipage}
	\vspace{-2ex}
\end{figure*}

\subsection{SUM BER Performance}

We use sum bit error rate (SUM BER) as metric to evaluate the decoding performance of proposed Jt-CDA decoder. Note that sum bit error is different from traditional bit error. For two devices, there are three sum bit outcomes (i.e., $\{0,1,2\}$), and any mismatch is treated as an error. SNR is defined as the reception power of superimposed signal versus noise.

In our simulation, the number of source bits is set to 1300. Given the total number of data subcarriers within an OFDM symbol of 48 (i.e., 24 bits for 1/2 convolutional codes), the total number of OFDM data symbols is 55. The constraint length of convolutional codes is set to 7, same in 802.11 standards. We compare the SUM BER performance of FSJD and RSJD for different SNRs (less than 20dB). For each SNR, we run 4000 times, and present the average result. For RSJD, the number of reduced-states is set to 128, 256 and 512 (the complexity of 128 states RSJD is the same with PSUD, since single-user decoder trellis has 64 states). We implement the whole OFDM system in Python and C, where C is used to accelerate the channel encoding and decoding.


We consider the following channel conditions (assuming unit single-path channel but with relative phase offset):
\begin{enumerate}
\item \textbf{Bad Channel}: Transmission signals pass through the AWGN channel and have no relative channel phase offset.
\item \textbf{Good Channel}: Transmission signals pass through the AWGN channel and have relative channel phase offset of $\pi/2$ radians.
\item \textbf{Near-realistic Channel}: Transmission signals pass through the AWGN channel and have random relative channel phase offset of $[0, 2\pi]$ radians, random time offset $[0,5]$ samples, random CFO $[-2,+2]$ kHz.
\end{enumerate}


Fig.~\ref{fig:phase0} shows the results of different Jt-CDA decoders in bad channel.
Given four possible constellation points, two of them are almost overlapped, which are hard to differentiate, and it may confuse the decoding. We can see that all Jt-CDA decoders show bad performance in this case, but joint decoders still outperform separate decoder. Since PSUD treats the other device as noise, it may lead to worse performance. 

Fig.~\ref{fig:phase90} shows the results of different Jt-CDA decoders in good channel. Four possible constellation points are located on the orthogonal coordinates and equally partition complex plane into four equal decision regions, leading to the minimal error rate.  We can see that FSJD outperforms RSJD since RSJD loses some path information during decoding. 

We also compare the performance of our Jt-CDA decoders under a near-realistic channel. We set random relative channel phase offset, random relative arrival time and random CFO in AWGN channel as mentioned above. Fig.~\ref{fig:phase_misalign} shows that FSJD and RSJD outperform PSUD 1-2dB when SNR$\geq$8dB. RSJD approaches FSJD when SNR$\geq$10 dB. To strike a balance between accuracy and complexity, we choose RSJD with $R$=512 as the Jt-CDA decoder in the simulations.

\subsection{Test Accuracy Performance}
The test accuracy is defined as the prediction accuracy of the global learned model on the test datasets. 
Different quantization bit lengths can result in different test accuracy on the same noise level. Long quantization bit length improves the test accuracy but also increases the data bits.
To achieve the optimal tradeoff, we first compare the test accuracy of four different quantization bit lengths. Fig.~\ref{quant} shows that $13$ is the optimal quantization bit length and we use that for the following simulations. Fig.~\ref{diff_noise} shows the test accuracy of digital AirComp based FEEL in 100 training iterations. We select RSJD as the Jt-CDA decoder and use the near-realistic channel mentioned above. We can see that: 1) different SNRs have different convergence rate; 2) the ideal noiseless condition gives the upper bound 72\% test accuracy; 3) the learning process may not converge in low SNR regime with test accuracy 10\% as the lower bound.

We also compare the final test accuracy between analog AirComp system \cite{zhu2019broadband} and our digital AirComp system under near-realistic channel. In analog AirComp system, the floating-point value parameters are directly loaded on subcarriers. We enhance \cite{zhu2019broadband} by allowing each subcarrier to transmit two parameters (in both I and Q planes). Since analog AirComp does not need quantization, it leads to less data transmission time. To make it fair, we allow analog AirComp to transmit the same data certain times so that the transmission duration is the same with digital AirComp (i.e., 13 times for 13-bit quantized digital AirComp), and pick the minimum mean squared error (MSE) transmission result as the final result. We also compute the test accuracy of the perfectly phase-aligned analog AirComp.
Fig.~\ref{real-simu} shows that: 1) digital AirComp can realize the same performance as the ideal (optimal) analog AirComp (around 70\%) in high SNR regime; 2) under phase asynchronous case, digital AirComp with Jt-CDA outperforms analog AirComp by at least 1.5 times when SNR$\geq$9dB; 3) digital AirComp with Jt-CDA outperforms digital AirComp with PSUD. All results show that digital AirComp is more suitable for realistic phase asynchronous systems.



\section{Conclusion}\label{conclusion}
In this paper, we present the first digital AirComp system for asynchronous OFDM-based FEEL systems. We design the digital AirComp transceiver architecture and a NOMA protocol. We also design different joint channel decoding and aggregation decoders. Simulation results demonstrate that digital AirComp can achieve accurate aggregation under phase asynchronous scenarios, while analog AirComp cannot achieve such goal even with high SNRs. In future, we plan to extend our digital AirComp system with more users and high-order QAM modulation. Another possible direction is to implement a real-time system on software-defined radio platforms.


\bibliography{IEEEabrv,bibfile}
\newpage
\ifCLASSOPTIONcaptionsoff
\newpage
\fi
\end{document}